\pdfoutput=1
\documentclass[11pt]{article}

\usepackage[final]{acl}

\usepackage{times}
\usepackage{latexsym}

\usepackage[T1]{fontenc}
\usepackage[utf8]{inputenc}

\usepackage{microtype}

\usepackage{inconsolata}

\usepackage{natbib}
\usepackage{multirow}
\setcitestyle{numbers,square}
\usepackage{algorithm}
\usepackage{hyperref}       % hyperlinks
\usepackage{url}            % simple URL typesetting
\usepackage{booktabs}       % professional-quality tables
\usepackage{amsfonts}       % blackboard math symbols
\usepackage{nicefrac}       % compact symbols for 1/2, etc.
\usepackage{xcolor}         % colors
\usepackage{color}

\usepackage{graphicx}
\usepackage{relsize}
\usepackage{amsfonts}
\usepackage{url}
\usepackage{graphicx}
\usepackage{subfigure}
\usepackage{caption}
\usepackage{hhline}
\usepackage{colortbl}
\usepackage{tabularx}
\usepackage{amsfonts}
\usepackage{amssymb}
\usepackage{amsthm,amsmath}
\usepackage{mathrsfs}
\usepackage{setspace}
\usepackage{lipsum}
\usepackage{listings}
\usepackage{multicol}
\usepackage{algpseudocode}% http://ctan.org/pkg/algorithmicx

\definecolor{mygrey}{RGB}{192, 192, 192} % 定义颜色
\definecolor{myred}{RGB}{216, 67, 47} % 定义颜色
\definecolor{mygreen}{RGB}{99, 218, 61} % 定义颜色

\title{StepCoder: Improve Code Generation \\ with Reinforcement Learning from Compiler Feedback}

\author{
    Shihan Dou$^{1}$\thanks{{ }\ Equal contributions.}$^{\dag}$, \ \  Yan Liu$^{1*}$, \ \ Haoxiang Jia$^{2}$, \ \ Limao Xiong$^{1}$, \ \ Enyu Zhou$^{1}$, Wei Shen$^{1}$, \\
    \textbf{Junjie Shan}$^3$\textbf{,} \ \ \textbf{Caishuang Huang}$^{1}$\textbf{,} \ \ \textbf{Xiao Wang}$^1$\textbf{,} \ \ \textbf{Xiaoran Fan}$^1$\textbf{,} \ \ \textbf{Zhiheng Xi}$^1$\textbf{,}\\
    \textbf{Yuhao Zhou}$^1$\textbf{,} \ \ \textbf{Tao Ji}$^1$\textbf{,} \ \ \textbf{Rui Zheng}$^{1\dag}$\textbf{,} \ \ \textbf{Qi Zhang}$^{1}$\thanks{{ }{ }Correspondence to: shdou21@m.fudan.edu.cn, \{rzhen g20, qz, tqui\}@fudan.edu.cn}\textbf{,} \ \ \textbf{Xuanjing Huang}$^1$\textbf{,} \ \ \textbf{Tao Gui}$^{1\dag}$ \\ 
    \normalsize{$^1$  Fudan NLP Lab,\ Fudan University,\ China} \\
    \normalsize{$^2$  Huazhong University of Science and Technology,\ China} \\
    \normalsize{$^3$  KTH Royal Institute of Technology,\ Sweden} \\
}

\begin{document}
\maketitle

\begin{abstract}

The advancement of large language models (LLMs) has significantly propelled the field of code generation.
Previous work integrated reinforcement learning (RL) with compiler feedback for exploring the output space of LLMs to enhance code generation quality.
However, the lengthy code generated by LLMs in response to complex human requirements makes RL exploration a challenge.
Also, since the unit tests may not cover the complicated code, optimizing LLMs by using these unexecuted code snippets is ineffective.
To tackle these challenges, we introduce \textbf{StepCoder}, a novel RL framework for code generation, consisting of two main components: CCCS addresses the exploration challenge by breaking the long sequences code generation task into a \textbf{C}urriculum of \textbf{C}ode \textbf{C}ompletion \textbf{S}ubtasks, while FGO only optimizes the model by masking the unexecuted code segments to provide \textbf{F}ine-\textbf{G}rained \textbf{O}ptimization.
In addition, we furthermore construct the APPS+ dataset for RL training, which is manually verified to ensure the correctness of unit tests.
Experimental results show that our method improves the ability to explore the output space and outperforms state-of-the-art approaches in corresponding benchmarks.
Our dataset APPS+ and StepCoder are available online \footnote{\url{https://github.com/Ablustrund/APPS\_Plus}}.

\end{abstract}

\section{Introduction}
Code generation or program synthesis aims to automatically generate source code that adheres to a specified programming requirement, which is typically described in natural language \cite{svyatkovskiy2020intellicode, gulwani2017program}.
% The code generated by models often exhibits considerable length and contains multiple conditional statements due to the variety of human requirements.
% Due to the complexity of human requirements, the code generated by models often exhibits considerable length and contains multiple conditional statements.
Recently, with the development of large language models (LLMs), techniques based on LLM \cite{li2023starcoder, touvron2023llama, luo2023wizardcoder} have demonstrated impressive ability in code generation.
% It is designed to boost the programming efficiency of developers and is gaining increasing attention from researchers due to its potential to automate programming in the future \cite{manna1971toward, kaddour2023challenges}.
However, challenges persist in aligning these models with complex human requirements \cite{austin2021program, apps, roziere2023codellama}, indicating a gap that still exists in fully meeting user expectations. 
% In this context, learning from human feedback \cite{ouyang2022training, zheng2023improving} exhibits impressive potential to improve the ability of these models to follow instructions more accurately and enhance dialogue quality \cite{bai2022training}.

In this context, learning from compiler feedback exhibits impressive potential to improve the comprehension of complicated human requirements and the quality of generated codes \cite{le2022coderl}.
This feedback from compilation and execution results is instrumental in directly ascertaining the functional correctness of programs \cite{li2022competition, wang2022compilable}.
Researchers \cite{liu2023rltf,shojaee2023execution} introduce reinforcement learning (RL) and leverage compiler feedback from unit tests as a reward metric to guide the exploration of the output space of LLMs.
The intention is for the policy model to favor actions that yield higher rewards increasingly.
Nevertheless, the optimization of LLMs for code generation via RL presents several hurdles.
\textbf{First}, the increasing complexity of human requirements often results in the generation of longer code sequences, which makes exploration struggle \cite{hao2023exploration, ladosz2022exploration}.
\textbf{Second}, in cases where a single unit test fails to cover the complex code, unexecuted code snippets may emerge that are not relevant to the reward. Rendering optimization based on the entire code sequence is potentially imprecise. 
% there exists the issue of incomplete code coverage by unit tests (i.e., code coverage < 100\%). 
% This limitation suggests that certain actions in the code sequence might not be pertinent to the reward \cite{malaiya1994relationship}, thereby rendering optimization based on the entire code sequence potentially imprecise. 
% Additionally, it is crucial to have accurate unit test samples that align with human requirements, ensuring that the feedback from the compiler is precise and reliable.
% However, we found that limitations in the quality of existing datasets such as APPS \cite{apps} for RL training.
% The absence of high-quality datasets exacerbates the difficulties in accurately learning from the compiler feedback through RL.
% Additionally, accurate unit test samples that match human requirements are essential for precise, reliable compiler feedback. 
% However, our analysis reveals quality limitations in existing datasets like APPS \cite{apps} for RL training, which impedes accurate learning from compiler feedback through RL.
Additionally, our analysis reveals quality limitations in existing datasets like APPS \cite{apps} for RL training, which impedes accurate learning from compiler feedback through RL.

To tackle these challenges, we first introduce StepCoder, an innovative framework developed for enhancing code generation through reinforcement learning. 
StepCoder integrates two key components: \textbf{C}urriculum of \textbf{C}ode \textbf{C}ompletion \textbf{S}ubtasks (CCCS) and \textbf{F}ine-\textbf{G}rained \textbf{O}ptimization (FGO). 
CCCS is designed to alleviate the complexities associated with exploration in code generation, while FGO is designed to provide more precise and effective optimization strategies.
Specifically, CCCS employs a step-by-step strategy to break down complex exploration problems (i.e., code generation) into a curriculum of easier sub-tasks (i.e., code completion).
As the training progresses, the difficulty of code completion tasks rises by increasing the portion of code that needs to be completed.
Eventually, the aim is for the model to evolve to a stage where it can effectively generate code solely from human requirements, thus fulfilling the original training goal of code generation.
On the other hand, the key insight of FGO is that code snippets that are not executed in a unit test do not contribute to the final reward calculation. 
Therefore, FGO uses a dynamic masking technique to mask unexecuted snippets from unit test evaluations, ensuring that the model is optimized utilizing only the relevant code segments.

Subsequently, our endeavor involves the development of APPS+, a dataset of superior quality specifically curated for code generation. 
APPS+ is meticulously designed to exclude code segments that exhibit syntax errors, are irrelevant to the stipulated problem, or fail to produce any output. 
Additionally, we have taken measures to standardize the format of inputs and outputs in unit tests to guarantee deterministic output comparisons.
% This dataset excludes generated code with syntax errors, code irrelevant to the given problem, or code producing no output. Furthermore, we format the inputs and outputs of unit tests, ensuring the output comparison was deterministic. 
% Finally, the APPS+ includes 7,456 problems with all generated code being executable.
% We then propose StepFiner, a novel framework for code generation via reinforcement learning.
% StepFiner introduces Stepper to reduce the difficulty of exploration in code generation and Finer to offer more precise optimization.

We evaluate the effectiveness of popular LLMs on APPS+.
The results reveal that although LLMs show progressive improvements, they face difficulties with complex human requirements.
We further evaluate our method on several extensively used benchmarks including MBPP \cite{austin2021program} and HumanEval \cite{chen2021evaluating}.
The experimental results show that StepCoder effectively eases the exploration difficulty in code generation, outperforming other reinforcement learning-based methods in effectiveness.
% Specifically, we introduced \emph{code scope}, a concept that defines code visibility, enabling the creation of more accurate sub-sequences. 
% We automatically collect the code scopes of canonical solutions for human requirements.
% Initially, the policy model is trained to complete code starting from scopes near a canonical solution's end.
% In other words, the model learns how to reach the goal and obtain rewards starting from states near the goal.
% The starting state progressively moves towards the first code scope, eventually making the model learn to generate code when only human requirements are provided (i.e., the original training objective of code generation).
% When exploring starting from states distant from the goal, the policy model can reach the goal by utilizing states that already know how to navigate to the goal.
The main contributions of our paper are as follows:
\begin{itemize}
\item We introduce StepCoder, a novelty training method via RL, including CCCS and FGO. 
CCCS makes exploration easier by breaking down the complicated goals into sub-objectives curriculum.
FGO provides fine-grained optimization by only utilizing the executed code in unit tests.
\item We constructed APPS+, a high-quality dataset designed for code generation. APPS+ provides a more rigorous evaluation of LLMs' capabilities and a foundation to introduce reinforcement learning in the training phase.
\item Experiments show that StepCoder can improve the exploration efficiency and effectiveness and outperform other methods.
\end{itemize}

\begin{figure}[t]
\centering
\includegraphics[width=0.46\textwidth]{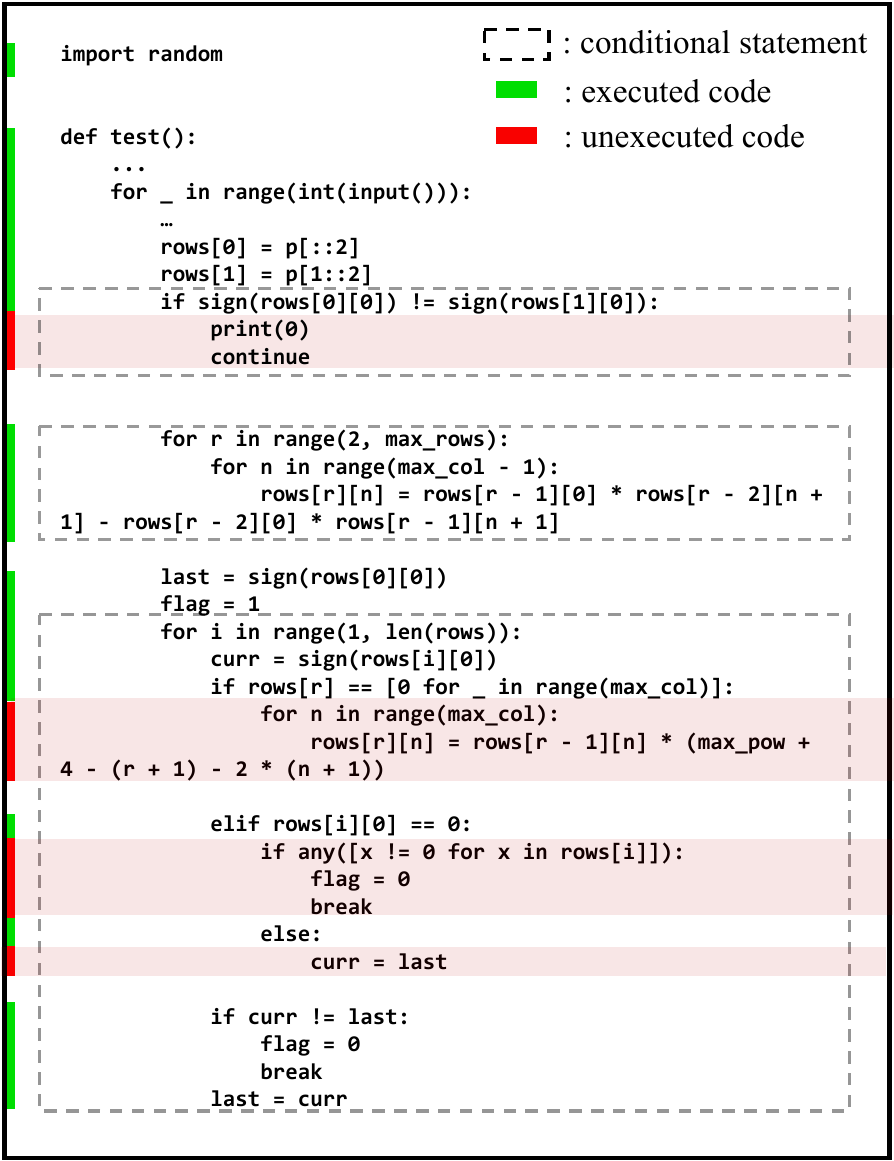}
\caption{The canonical solution of an instance in the APPS dataset.
We collect the conditional statements by analyzing their abstract syntax tree, and some conditional statements are highlighted with a grey dashed box.
When inputting $s=[1\text{\textbackslash n}10\,12\,1\,5\,3\text{\textbackslash n}]$, only 75\% of the code fragment is executed, highlighted with a green background.
}
\vspace{-1em}
\label{fig:an-instance}
\end{figure}

\section{Motivation}
\label{sec:motivation}
% To understand the key purpose behind creating the APPS+ dataset, we performed an analysis of the popular dataset APPS \cite{apps}, which was widely used for RL training in code generation.
% Additionally, 
In this section, we clearly illustrate the challenges faced by reinforcement learning in code generation using a simplified example from APPS \cite{apps}, which was widely used for RL training in code generation.

% \textbf{A brief analysis of the APPS dataset.}
% Reinforcement learning requires an amount of high-quality training data. During our investigation, we found that among the currently available open-source datasets, only APPS meets this requirement.
% However, we found plenty of incorrect instances in APPS. 
% Our analysis revealed issues in over 1,500 instances, such as missing or inconsistent input and output data. 189 instances had uncompilable code, and over 1,000 failed to execute correctly. Additionally, the outputs of canonical solutions in over 500 instances did not match the expected results, undermining the benchmark's reliability.

% Further investigation showed that compile failures were due to incorrect Python syntax and incomplete canonical code. Execution issues were often caused by missing dependencies or API misuse. The mismatch in expected outputs was typically due to incorrect output or discrepancies between standard code and the programming problems.

\textbf{Exploration problems of RL in code generation.}
Exploration methods play a crucial role in tackling complicated sequence but sparse reward problems \cite{yang2021exploration, ladosz2022exploration}.
When a policy model explores a trajectory with high returns, it undergoes optimization, making it inclined to take similar actions in the future \cite{williams1992simple, salimans2018learning}.
% Unfortunately, in complex tasks with sparse rewards, high-return trajectories are difficult to explore \cite{sutton1999policy}.

Consider the code shown in Figure~\ref{fig:an-instance}, aimed at fulfilling a given human requirement.
We first collect the conditional statements (CS) that are indicated by the dashed box by analyzing its abstract syntax tree. 
Conditional statement introduces new independent paths, increasing the complexity of the program \cite{shepperd1988critique}.
Suppose $P_\theta(\text{CS}_i)$ denotes the probability that the policy model with parameter $\theta$ completes the $i$-th conditional statement.
The probability that the policy model correctly generates this code according to human requirements can be expressed as follows:
\begin{equation}
\small
    P \propto P_o \prod \limits_{i=1}^3 P_\theta(\text{CS}_i),
\end{equation}
where $P_o$ is the probability of other code snippets except the code labeled in the figure.
Typically, we initialize the policy model with the SFT model in sequence generation tasks to facilitate easier exploration \cite{ouyang2022training, zheng2023delve}.
However, the limited performance of the SFT model in code generation still leads to the probability $P_\theta(\text{CS}_i)$ at low values \cite{shojaee2023execution, roziere2023codellama}.
The increasing complexity of human requirements in code generation tasks often leads to a corresponding rise in the number of conditional statements. This escalation can result in a substantial decrease in the probability $P_\theta(\text{CS}_i)$, potentially leading $P$ to an exponential reduction. 
Such a scenario exacerbates the challenges associated with exploration in large language models.
An alternative approach to facilitate exploration is through reward shaping, a technique where designers \emph{artificially} introduce rewards more frequently \cite{ladosz2022exploration}. 
However, this method encounters a significant limitation in the context of our application. 
Specifically, in code generation tasks utilizing unit test feedback, rewards can only be obtained after the execution of the completely generated code.
Consequently, the exploration of high-return trajectories in tasks with complex sequences and sparse rewards poses a significant challenge in optimizing the policy model.
% In other words, high-return trajectories are difficult to explore in this complicated sequence but sparse reward task, making optimizing the policy model challenging.

\textbf{Optimization problems of RL in code generation.}
We first introduce the RL fine-tuning process in code generation.
Formally, for a learned policy model $\pi_\theta$ with parameter $\theta$, we treat the prediction of each token as an \emph{action} $a$ taken by $\pi_\theta$ according to the history token sequences.
The history token sequences can be viewed as the \emph{state} $s$.
Given a human requirement $x$, we denote the solution code $y$ generated by $\pi_\theta$ as an episode, and $r(x,y)$ is the reward function from the compiler based on compilation and execution.
Updating the parameters of $\pi_\theta$ by using gradient policy algorithm \cite{sutton1999policy} can be represented as follows:
\begin{equation}
\label{eq:motivation-pg}
% \small
    \max_\theta \ E_{(x,y) \sim D_{\pi_\theta}} [\sum_t A_{\pi}^{t} \log (y_t|y_{1:t-1}, x;\theta) ]
\end{equation}
where $A_\pi$ is the advantage computed by the Generalized Advantage Estimator (GAE) \cite{schulman2015high} from reward $r$, to reduce the variability of predictions.

In code generation, rewards are contingent upon the correctness of the unit test sample, which is only relevant to the code snippet being executed.
For instance, as shown in Figure~\ref{fig:an-instance}, when the input to the function is $[1\text{\textbackslash n}10\,12\,1\,5\,3\text{\textbackslash n}]$, 75\% of the code fragment is executed, which is highlighted with a green dashed box.
It indicates that some actions in the code are irrelevant to the reward, which leads to inaccurate advantage. Therefore, optimizing the policy model $\pi_\theta$ with all actions is ineffective by using Equation~\ref{eq:motivation-pg}.

\begin{figure*}[t]
\centering
\includegraphics[width=0.96\textwidth]{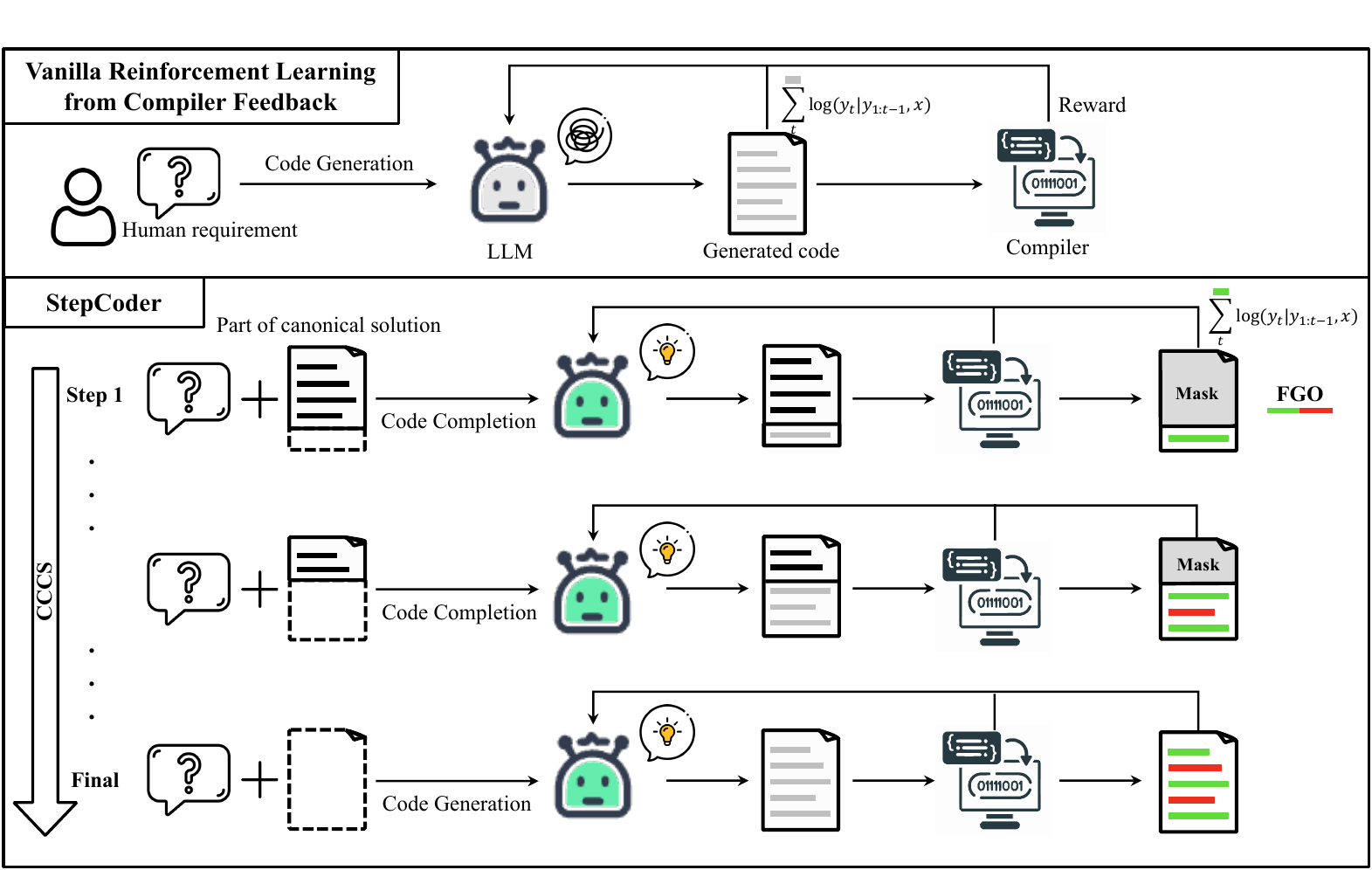}
\caption{The overview of our method.
In code generation, the environment with sparse and delayed rewards and the complicated human requirement that involves a long sequence make exploration challenging for the Vanilla RL.
In CCCS, we break down a complicated exploration problem into a curriculum of sub-tasks. 
Utilizing a portion of the canonical solution as the prompt enables the LLM to explore starting from simple sequences.
The computation of rewards is only relevant for the executed code snippets, and it is imprecise to optimize the LLM with the entire code (i.e., \textcolor{mygrey}{\rule{0.2cm}{0.2cm}}).
In FGO, we mask unexecuted tokens (i.e., \textcolor{myred}{\rule{0.2cm}{0.2cm}}) in unit tests and only compute the loss function using executed tokens (i.e., \textcolor{mygreen}{\rule{0.2cm}{0.2cm}}) to provide a fine-grained optimization.
}
\vspace{-1em}
\label{fig:fig-main}
\end{figure*}

\section{Method}
In this section, we elaborate on the methodological details of StepCoder, which provide an easier exploration and fine-grained optimization for RL in code generation, respectively, as shown in Figure~\ref{fig:fig-main}.

\subsection{Priliminaries}
Suppose $\mathcal{D} = \{(x_i, y_i, u_i, e_i)\}_{i=0}^{N}$ is the training dataset for code generation, which $x$, $y$, $u$ denotes the human requirement (i.e., the task description), the canonical solution and the unit test samples, respectively.
$e_i = \{st_j, en_j\}_{j=0}^{E_i}$ is a list of conditional statements by automatically analyzing the abstract syntax tree of the canonical solution $y_i$, which $st$ and $en$ represent the start position and the end position of the statements, respectively.
$e$ is sorted in ascending order based on the start position $st$.
For a human requirement $x$, its canonical solution $y$ can be represented as $\{a_t\}_{t=0}^{T}$.
In code generation, given a human requirement $x$, the final states are the set of codes passing the unit tests $u$.

\subsection{StepCoder}
StepCoder integrates two key components: CCCS and FGO.
CCCS is designed to break the code generation tasks into a curriculum of the code completion subtasks.
It can alleviate the exploration challenge in RL.
FGO is specifically designed for code generation tasks to provide fine-grained optimization by computing only the loss of executed code snippets.

\textbf{CCCS.}
In code generation, the solution to a complicated human requirement usually involves a long action sequence taken by the policy model.
Meanwhile, the feedback from the compiler is delayed and sparse, i.e., the policy model only receives the reward after generating the entire code.
In this scenario, exploring is difficult.
The core of our method is to break down such a long sequence of exploration problems into a curriculum of short, easily explorable sub-tasks.
We simplify code generation to code completion sub-tasks.
These sub-tasks are automatically constructed from the canonical solution in the training dataset.

Consider a human requirement $x$, early in the training phase of CCCS, the starting point $s^*$ of exploration is the states near the final states.
Specifically, we provide the human requirement $x$ and the front part of the canonical solution $x_p = \{a_i\}_{i=0}^{s^*}$, and the policy model is trained to complete the code based on $x^{'} = (x, x_p)$.
Let $\hat{y}$ be the combined sequence of $x_p$ and the output trajectory $\tau$, i.e. $\hat{y} = (x_p, \tau$).
The reward model provides the reward $r$ according to the correctness of the code snippet $\tau$ with $\hat{y}$ as input, where we use the same setting as previous approaches \cite{le2022coderl, shojaee2023execution} as follows:
\begin{align}
% \small
\label{eq:reward}
r(x^{'},\hat{y}) = 
\begin{cases}
~+1,~ \text{if $\hat{y}$ passed all unit tests}\\
-0.3,~ \text{if $\hat{y}$ failed any unit test}\\
-0.6,~ \text{if $\hat{y}$ happened runtime error}\\
~-1,~ \text{if $\hat{y}$ happened compile error}.
\end{cases}
\end{align}
We use the Proximal Policy Optimization (PPO) algorithm \cite{schulman2017proximal} to optimize the policy model $\pi_\theta$ by utilizing the reward $r$ and the trajectory $\tau$.
In the optimization phase, the canonical solution's code segment $x_p$ used for providing prompts is masked, such that it does not contribute to the gradient for the policy model $\pi_\theta$ update.
CCCS optimizes the policy model $\pi_\theta$ by maximizing the objection function as follows:
\begin{align}
\label{eq:vanilla-rl}
\small
    \text{Objective}(\theta) &= E_{(x^{'}, \hat{y}) \sim D_{\pi_\theta}} [r(x^{'}, \hat{y}) & \nonumber \\
    &- \beta \log (\pi_\theta (\hat{y} | x^{'})) / \pi^{\text{ref}} (\hat{y} | x^{'}) ]
\end{align}
where $\pi^{\text{ref}}$ is the reference model in PPO, which is initialized by the SFT model.

As the training progresses, the starting point $s^*$ of exploration gradually moves towards the beginning of the canonical solution.
Specifically, we set a threshold $\rho$ for each training sample.
Each time the cumulative correct proportion of code segments generated by $\pi_\theta$ is greater than $\rho$, we move the starting point toward the beginning.
In the later stages of training, the exploration of our method is equivalent to the exploration process of original reinforcement learning, i.e., $s^* = 0$, where the policy model generates code using only human requirements as input.

The starting point $s^*$ is sampled at the beginning position of the conditional statements to complete the remaining unwritten code segments.
Specifically, a program with a greater number of conditional statements results in increased independent paths, leading to a higher logical complexity \cite{shepperd1988critique}. 
This complexity necessitates more frequent sampling to improve the quality of training, while programs with fewer conditional statements need less frequent sampling.
This sampling method allows for a balanced and representative sampling of code structures, catering to both complex and simple semantic constructs in the training dataset.
To accelerate the training phase, we set the $i$-th sample's number of curricula equal to $\lceil \sqrt{E_i} \rceil$, where $E_i$ is its number of conditional statements.
The $i$-th sample's stride of the training curriculum is $\lceil \frac{E_i}{\lceil \sqrt{E_i} \rceil} \rceil$ instead of one.

The key insight of CCCS can be summarized as follows: 
1) It is easy to explore from the states near the goal (i.e., final states).
2) Exploring starting from the states distant from the goal is challenging, but it becomes easier when can leverage states that have already learned how to reach the goal.

\textbf{FGO.}
The relationship between reward and action in code generation differs from other reinforcement learning tasks such as Atari \cite{mnih2015human, lillicrap2015continuous}.
In code generation, we can exclude a set of actions irrelevant to computing the rewards in generated code.
Specifically, as mentioned in Section~\ref{sec:motivation}, for a unit test, the feedback from the compiler relates only to the code snippets being executed.
However, in vanilla RL optimization objectives, as shown in Equation~\ref{eq:vanilla-rl}, all actions of the trajectory are engaged in the computation of the gradient used in the policy update, which is imprecise.

To improve the precision of optimization, we mask actions (i.e., tokens) that are not executed in unit tests when computing the loss for updating the policy model.
The full algorithm of CCCS and FGO is detailed in Algorithm~\ref{alg:algorithm1}.

\section{Experiments}
In this section, we first introduce APPS+, a high-quality dataset for code generation by manually verifying based on the APPS dataset.
Then, we elaborate on the experiment details and the experimental results.

\begin{table*}[htbp]
% \small
  \centering
  \begin{spacing}{0.85}
    \setlength{\tabcolsep}{2.2mm}{%7可随机设置，调整到适合自己的大小为止
% Table generated by Excel2LaTeX from sheet 'Sheet2'
\begin{tabular}{lc|ccc|c}
\toprule
\toprule
\multicolumn{1}{c}{\multirow{2}[2]{*}{\textbf{Models}}} & \multicolumn{1}{c|}{\multirow{2}[2]{*}{\textbf{Size}}} & \multicolumn{4}{c}{\textbf{APPS+}} \\
      &       & \multicolumn{1}{c}{\textbf{Introductory}} & \multicolumn{1}{c}{\textbf{Interview}} & \multicolumn{1}{c|}{\textbf{Competition}} & \multicolumn{1}{c}{\textbf{Overall}} \\
\midrule
\multicolumn{6}{c}{Base Models} \\
\midrule
CodeLlama \cite{roziere2023codellama} & 13B    & 18.7     & 11.0     & 0.0     & 13.0 \\
CodeLlama-Python \cite{roziere2023codellama} & 13B    & 29.0     & 12.3     & 2.9     & 17.9 \\
DeepSeek-Coder-Base \cite{Guo2024DeepSeekCoderWT} & 6.7B    & 13.0     & 10.3     & 5.0     & 10.9 \\
\midrule
\multicolumn{6}{c}{Supervised Fine-tuned Models} \\
\midrule
StarCoder \cite{li2023starcoder} & 15.6B    & 6.3     & 4.1     & 0.7     &  4.7 \\
CodeLlama-Instruct \cite{roziere2023codellama} & 13B    & 33.3     & 11.0     & 1.4     & 18.7 \\
WizardCoder-Python-V1.0 \cite{luo2023wizardcoder} & 13B    & 39.7     & 15.1     & 4.3     & 23.6 \\
DeepSeek-Coder-Instruct \cite{Guo2024DeepSeekCoderWT} & 6.7B    & 49.4     & 18.7     & 3.6     & 29.2 \\
SFT on APPS+ & 6.7B    & \textbf{50.1}     & \textbf{19.0}   & \textbf{6.4}   & \textbf{29.8} \\
\midrule
\multicolumn{6}{c}{Reinforcement Learning-based Models (Using DeepSeek-Coder-Instruct-6.7B as the backbone)} \\
\midrule
Vanilla PPO & 6.7B    & 53.7     & 20.1     & 5.0     & 31.7 \\
PPOCoder \cite{shojaee2023execution} & 6.7B    & 54.4     & 20.3     & 6.4     & 32.1 \\
RLTF \cite{liu2023rltf} & 6.7B    & 55.1     & 20.8     & 6.4     & 32.7 \\
\midrule
\textbf{StepCoder (Ours)} & 6.7B    & \textbf{59.7}     & \textbf{23.5}     & \textbf{8.6}     & \textbf{36.1} \\
\quad \textbf{w/o} CCCS & 6.7B    & 58.7     & 21.7     & 7.1     & 34.6 \\
\quad \textbf{w/o} FGO & 6.7B    & 58.4     & 23.3     & 8.6     & 35.5 \\
\bottomrule
\bottomrule
\end{tabular} }%
\end{spacing}
\caption{Results of pass@1 on our proposed APPS+. We compare popular and widely used state-of-the-art methods with our method.
To ensure a fair comparison, we apply these RL-based methods using the same base model (i.e., DeepSeek-Coder-Instruct-6.7B \cite{Guo2024DeepSeekCoderWT}) as a backbone on the APPS+ dataset.
In addition, We conduct supervised fine-tuning using our APPS+ dataset based on DeepSeek-Coder-Instruct-6.7B to further validate the effectiveness and necessity of our approach. }
\label{tab:main-results}%
\vspace{-1.5em}
\end{table*}%

\subsection{Dataset Preprocessing}
Reinforcement learning requires an amount of high-quality training data. During our investigation, we found that among the currently available open-source datasets, only APPS meets this requirement.
However, we found there are incorrect instances, such as missing input, output, or canonical solution, canonical solutions that were uncompileable or unexecutable, and discrepancies in execution output. 
% Our analysis revealed issues in over 1,500 instances, such as missing or inconsistent input and output data. 189 instances had uncompilable code, and over 1,000 failed to execute correctly. Additionally, the outputs of canonical solutions in over 500 instances did not match the expected results, undermining the benchmark's reliability.

% Further investigation showed that compile failures were due to incorrect Python syntax and incomplete canonical code. Execution issues were often caused by missing dependencies or API misuse. The mismatch in expected outputs was typically due to incorrect output or discrepancies between standard code and the programming problems.
To refine the APPS dataset, we excluded instances lacking input, output, or canonical solutions. Then, we standardized the formats of input and output to facilitate the execution and comparison of unit tests. We conducted unit tests and manual analysis for each instance, eliminating those with incomplete or irrelevant code, syntax errors, API misuse, or missing library dependencies. For discrepancies in output, we manually reviewed the problem description, correcting the expected output or eliminating the instance.

Finally, we construct the APPS+ dataset, containing 7,456 instances. Each instance includes a programming problem description, a canonical solution, a function name, unit tests (i.e., inputs and outputs), and starter code (i.e., the beginning part of the canonical solution). 
Appendix~\ref{sec:appendix-instance} illustrates an example from APPS+. The top section of the figure shows the problem description, and the right section presents the canonical solution, unit tests, and metadata.  Further details of APPS+ are discussed in Appendix~\ref{sec:appendix-exp-benchmarks}.

\subsection{Experiment Details}

\textbf{Benchmarks.}
In our study, we initially evaluated our method and baselines on our pre-processed \textbf{APPS+} dataset.
Moreover, we also evaluate these methods on several widely-used benchmarks in code generation, i.e., \textbf{MBPP} (Mostly Basic Programming Problems) \cite{austin2021program} and \textbf{HumanEval} \cite{chen2021evaluating}.
We evaluate the MBPP and HumanEval benchmark in a zero-shot learning setting which is the same as previous approaches \cite{le2022coderl, shojaee2023execution}.
In this setting, we fine-tune the models only on the APPS+ dataset and evaluate the code generation performance on MBPP and HumanEval.
The detailed description of benchmarks can be found in the Appendix~\ref{sec:appendix-exp-benchmarks}.

\textbf{Baselines.}
To verify the effectiveness of StepCoder and evaluate the performance of LLMs on our APPS+ dataset, we consider a wide range of baselines, including StarCoder \cite{li2023starcoder}, WizardCoder \cite{luo2023wizardcoder}, DeepSeek-Coder \cite{Guo2024DeepSeekCoderWT}, and three versions of CodeLlama (Base, Python, Instruct) \cite{roziere2023codellama}.
Moreover, we also consider vanilla PPO and two state-of-the-art RL-based approaches, including PPOCoder \cite{shojaee2023execution} and RLTF \cite{liu2023rltf}.
We carried out experiments applying these methods utilizing the same backbone (i.e., DeepSeek-Coder-Instruct \cite{Guo2024DeepSeekCoderWT}) on the APPS+ dataset to ensure a fair comparison.
In addition to demonstrating the necessity and effectiveness of our method, we also supervised fine-tuning DeepSeek-Coder-Instruct \cite{Guo2024DeepSeekCoderWT} on the APPS+ dataset to exclude the effect of training data.
The detailed description of these baselines is discussed in Appendix~\ref{sec:appendix-exp-baselines}.

\textbf{Implementation Details.} 
During the SFT phase, we adopt a learning rate set at $2e^{-5}$, conduct training for three epochs, and employ a warm-up period of $0.3$ epochs, with a linear decay to zero.
The fine-tuning process was conducted on a device with eight NVIDIA A100 80G GPUs, with the global batch size set to $64$.
In the PPO training phase, we employ a learning rate of $5e^{-7}$ for the policy model and $1.5e^{-6}$ for the critic model.
For each example, we collect a $16$ roll-out code using nucleus sampling. 
The sampling temperature is set to $0.8$, top-p is set to $0.9$, and the maximum output token length is set to $1024$. 
The token-level KL penalty coefficient $\beta$ is set to $0.05$, with a clip value of $0.8$. 
In the decoding phase, the temperature and top\_p are set to $0.2$ and $0.95$, respectively.

\textbf{Evaluation \& Metric.}
We conduct the experiments based on Python3.x.
Note that we also use Python3.x during the reward collection in RL-based methods.
Following prior studies \cite{roziere2023codellama, luo2023wizardcoder, le2022coderl}, we use \textbf{Pass@k} \cite{chen2021evaluating} metric to evaluate all the models.
Pass@k quantifies the proportion of instances in which at least one of the k-generated code solutions per human requirement successfully passes all unit tests.
The prompts used for code generation are listed in Appendix~\ref{prompt}.

\subsection{Experimental Results on APPS+}
To assess the performance of widely used LLMs and our StepCoder on code generation, we conduct experiments on the APPS+ dataset that we constructed.
The experimental results are illustrated in Table~\ref{tab:main-results}.
The results indicate that RL-based models outperform other language models, including both base models and SFT models.
It is reasonable to infer that reinforcement learning can further enhance the quality of code generation by more effectively navigating the model's output space, guided by compiler feedback.

Furthermore, our StepCoder surpasses all baseline models including other RL-based approaches, achieving the highest score.
Specifically, our approach obtains 59.7\%, 23.5\% and 8.6\% in the `Introductory', `Interview' and `Competition', respectively.
Our approach excels in exploring the output space compared to other RL-based methods, achieved by simplifying complex code generation tasks to code completion sub-tasks. 
Additionally, the FGO process plays a pivotal role in precisely optimizing the policy model.
We also found that the performance of StepCoder is better than LLM which supervised fine-tuning on the APPS+ dataset based on the same backbone.
The latter did little to improve the pass rate of the generated code compared with the backbone.
This also directly demonstrates that the method of using compiler feedback to optimize the model improves the quality of the generated code better than next-token prediction in code generation.

\begin{table}[htbp]
% \small
  \centering
  \begin{spacing}{0.9}
    \setlength{\tabcolsep}{0.5mm}{%7可随机设置，调整到适合自己的大小为止
% Table generated by Excel2LaTeX from sheet 'Sheet2'
\begin{tabular}{lc|c}
\toprule
\toprule
\multicolumn{1}{c}{\textbf{Models (6.7B)}} & \textbf{HumanEval} & \textbf{MBPP} \\
\midrule
DeepSeek-Coder-Instruct &   \textbf{78.0}  & \textbf{64.2} \\
SFT on APPS+ & 55.5  & 54.8    \\
\midrule
Vanilla PPO &   78.0   & 65.0 \\
PPOCoder  & 76.8     & 63.8 \\
RLTF  & 76.8    & 65.2 \\
\textbf{StepCoder (Ours)} & \textbf{78.7}   & \textbf{67.0}  \\
\bottomrule
\bottomrule
\end{tabular} }%
\end{spacing}
\caption{
Results of pass@1 on MBPP and HumanEval. 
We evaluate the LLMs' performance on code generation in a zero-shot learning setting.
In this setting, the models are fine-tuned on our proposed APPS+ dataset and tested for their ability on MBPP and HumanEval.
}
\label{tab:results-mbpp-humaneval}%
\vspace{-1em}
\end{table}%

\subsection{Ablation Studies}

To investigate the impact of individual components in StepCoder, we conducted ablation experiments with two variations of our approach, including StepCoder only with CCCS and only with FGO.
The experimental results are presented in Table~\ref{tab:main-results}.
Experimental results demonstrate that both components of our approach improve the quality of the generated code compared to vanilla PPO.
CCCS can enhance its performance in addressing Competition-level problems.
This improvement is logical, considering that CCCS effectively simplifies the exploration of more complex human requirements.
Simultaneously, FGO boosts the pass rate of unit tests by integrating compiler feedback with the relevant executed code snippet.

\begin{figure}[htbp]
\centering
\includegraphics[width=0.42\textwidth]{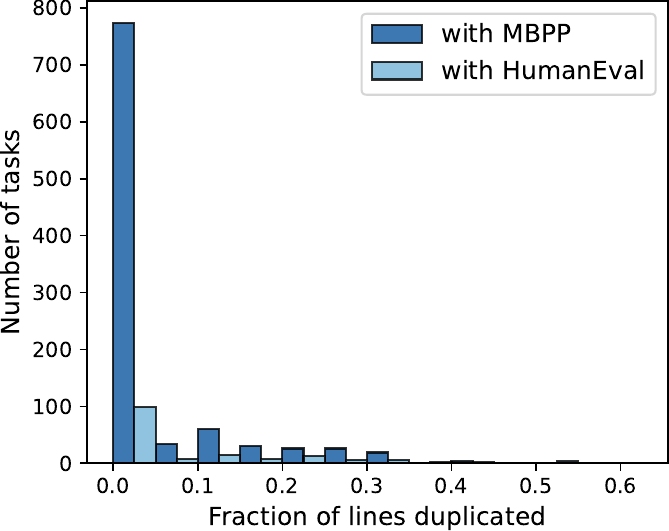}
\caption{
Analysis of duplicated lines between APPS+ and the two benchmarks.
The overlap of data between APPS+ and them is very small.
Only 0.2\% and 7.1\% had more than half of their lines matched somewhere in MBPP and HumanEval, respectively.
}
\vspace{-1em}
\label{fig:fig-overlap}
\end{figure}

\subsection{Results on MBPP and HumanEval}
To further demonstrate the effectiveness of our method, we conducted comparative analyses of StepCoder against various approaches using the well-recognized benchmarks MBPP and HumanEval.
These models are trained on APPS+ and then evaluated on MBPP and HumanEval.
The experimental results are illustrated in Table~\ref{tab:results-mbpp-humaneval} which shows that StepCoder is superior over all other models on both benchmarks.

However, there are concerns regarding potential overlaps in the training data between APPS+ and the two benchmarks, which might contribute to an improvement in performance.
To address these concerns, we analyze the difference between APPS+ and the benchmarks by calculating the code line overlap ratio of two corresponding canonical solutions following previous work \cite{austin2021program, le2022coderl}.
The findings are presented in Figure~\ref{fig:fig-overlap}.
This evidence underscores our approach's effectiveness in enhancing the quality of generated code and its capability across a broad spectrum of code generation tasks, primarily by improving the exploration problem in reinforcement learning.

Meanwhile, our findings revealed a significant degradation in the performance of the SFT model on both MBPP and HumanEval benchmarks.
Further analysis of the error cases showed that a minority were related to function name errors, while the majority were associated with program correctness errors.
This also indicated that SFT on a single dataset may impair the ability to follow instructions and the ability to generalize, thus affecting the performance of code generation on other tasks.
In contrast, RL-based methods can improve the performance for unseen tasks of code generation.

\subsection{Analysis by Unit Test Results}
\begin{figure}[htbp]
\centering
\includegraphics[width=0.46\textwidth]{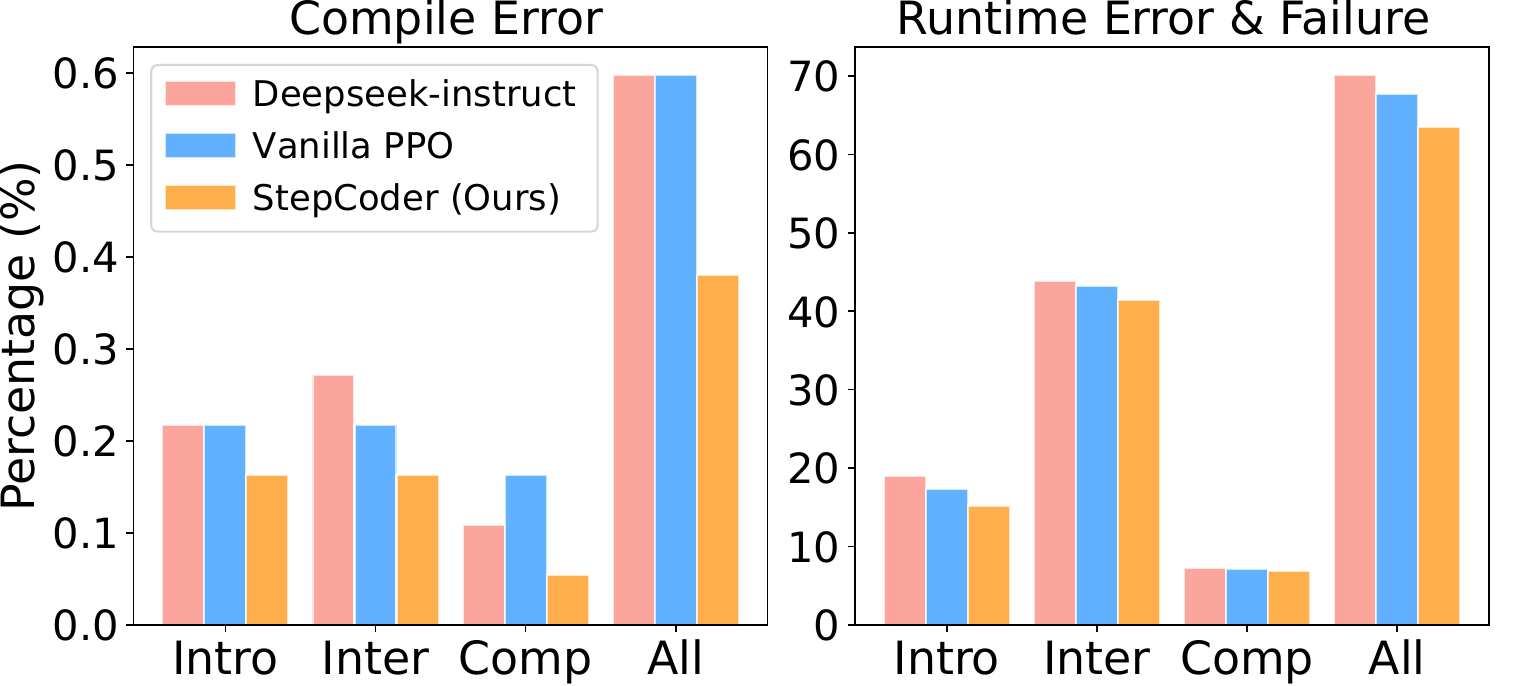}
\caption{
Analysis by unit test results on APPS+.
The results are categorized into CompileError (Reward = -1) and Runtimeerror \& Failure (Reward = -0.6 or -0.3).
}
\vspace{-1.8em}
\label{fig:fig-analysis}
\end{figure}

We further analyzed the results of cases that did not pass all unit tests, as shown in Figure~\ref{fig:fig-analysis}.
% The results show that our proposed method can reduce the probability of occurring compilation errors.
The results show that our proposed method can effectively reduce the likelihood of compilation errors, which is particularly evident in Interview-level and Competition-level programming problems.
% On the other hand, the probability of all LLMs occurring runtime errors and failures is substantially higher than that of compilation errors.
However, it was also observed that all LLMs are more prone to runtime errors and failures as compared to compilation errors, albeit StepCoder shows a comparatively lower rate of runtime errors and failures.
% StepCoder also shows a reduced rate of Runtime Error and Failure.
These results demonstrate that StepCoder is less prone to compilation errors, but still suffers from runtime errors and failure. 
Consequently, these findings suggest that future research should further concentrate on significantly reducing runtime errors, which could greatly enhance both the quality and the pass rate of the code generated by such models.

\section{Related Work}

\subsection{Large Language Models for Code Generation}
Recently, LLMs have shown remarkable ability in understanding natural language and code generation by training on large text corpora containing code data.
% There is increasing interest in employing large language models to generate code based on human requirements.
Several pre-trained language models (PLMs) demonstrate significant potential for code generation including CodeGPT \cite{lu2021codexglue}, PanGu-Coder \cite{christopoulou2022pangu}, SantaCoder \cite{allal2023santacoder}, CodeGeex \cite{zheng2023codegeex} and Phi-1.5 \cite{li2023textbooks2}.
% Phi-1.5 \cite{li2023textbooks2} further enhances the performance of code generation by improving the quality of the pre-trained training corpus.
% However, since these PLMs are not fine-tuned to align with instruction datasets, their capacity to follow human instructions on code generation remains limited.
In addition, SFT models achieve more competitive performance such as CodeX \cite{chen2021evaluating}, StarCoder \cite{li2023starcoder}, WizardCoder \cite{luo2023wizardcoder}, Code Llama Instruct \cite{roziere2023codellama}, and DeepSeek-Coder \cite{Guo2024DeepSeekCoderWT}.
% Additionally, WizardCoder \cite{luo2023wizardcoder} has proven to be highly effective in code generation by fine-tuning more complex instruction data.
% This complicated dataset is constructed by adapting the Evol-Instruct \cite{xu2023wizardlm} on code-related tasks.
% Nevertheless, these models heavily rely on high-quality datasets to avoid fitting incorrect code data and ignore important signals of code syntactic and functional correctness from unit tests \cite{chen2021evaluating, shojaee2023execution, liu2023rltf}.

Reinforcement Learning is a method of learning the optimal policy by exploring the environment and obtaining rewards \cite{williams1992simple, sutton1998introduction}.
Recently, some researchers have introduced RL to LLMs and improved the quality of the generated code by utilizing the unit test feedback to explore the output space of the policy model \cite{shojaee2023execution, liu2023rltf, le2022coderl}.
For instance, CodeRL \cite{le2022coderl} leverages signal from unit tests as rewards and utilizes the actor-critic approach \cite{konda1999actor, sutton1999policy} to enhance models on code generation.
% PPOCoder \cite{shojaee2023execution} refines CodeRL by employing the PPO algorithm \cite{schulman2017proximal} and computing rewards in more ways such as abstract syntax tree.
PPOCoder \cite{shojaee2023execution} refines CodeRL by employing the PPO algorithm \cite{schulman2017proximal} and RLTF \cite{liu2023rltf} provides fine-grained rewards through the error locations, but the reward space is still sparse.
However, the exploration of complex tasks in an environment characterized by a sparse reward is challenging.
These methods still fall short of effectively using RL to enhance the model's performance in code generation.

\subsection{Exploration in Reinforcement Learning}
Exploration is crucial in addressing long sequences and sparse reward problems \cite{hao2023exploration, ladosz2022exploration}.
In the sequence generation task, researchers improved exploration by initializing the policy model using the SFT model \cite{ouyang2022training, shen2023loose}. 
% Similar methods have also been incorporated into our proposed approach. 
% However, despite this method, challenges in exploration persist, necessitating additional methods to ensure effective exploration. 
Our proposed approach incorporates similar methods, but additional methods are necessary to ensure effective exploration.
This is particularly evident when facing complex human requirements, where the limited quality of code generated by SFT models makes exploration still challenging \cite{shojaee2023execution}.

Other notable methods introduce the Process-Supervised Reward Model to provide step-by-step rewards for complex sequence generation tasks such as mathematical reasoning and code generation \cite{uesato2022solving, lightman2023let, luo2023wizardmath, ma2023let}.
However, these methods require labelling a large preference dataset to train the reward model.
Similar to our approach, some methods construct a learning curriculum by initiating each episode from a sequence of progressively more challenging starting states \cite{salimans2018learning, hosu2016playing, florensa2017reverse}.
In contrast to our approach, these methods are designed to address the problem of exploration in other fields, such as gaming and robotic manipulation.
Meanwhile, our approach combines software engineering features to dynamically determine the starting states through conditional statements. 
We also introduce FGO to provide a fine-grained optimization for the policy model by leveraging the coverage information.

% \subsection{Benchmarks for Code Generation}
% The training and evaluation of code generation are highly dependent on the quality of the benchmarks. 
% Recently, researchers have delved into constructing datasets in different programming languages for various contexts \cite{lai2023ds, apps, austin2021program, chen2021evaluating}. 
% APPS \cite{apps} is one of the most widely used Python-focused benchmarks. 
% It contains 10,000 programming problems collected from public programming competition platforms. 
% HumanEval \cite{chen2021evaluating}, proposed by OpenAI, contains 164 human-written problems with a function signature, a reference code, and several unit tests. 
% Another benchmark, MBPP \cite{austin2021program}, consists of around 974 crowd-sourced Python programming problems, covering programming fundamentals and standard library functionality.
% However, the existing dataset is unsuitable as an RL training dataset, due to its limited quantity or quality.

\section{Conclusion}
In this paper, we introduce StepCoder, a novelty training framework via RL.
StepCoder breaks down complicated exploration problems to reduce the difficulty of exploring environments with sparse rewards while providing fine-grained optimization.
In addition, we also construct a high-quality dataset APPS+, specifically for code generation.
% to provide a more rigorous evaluation of LLM and a foundation to introduce RL during the training phase.
Experiments indicate that our method can effectively improve the quality of generated code via reinforcement learning compared to other approaches.

% \section{Limitations}

\bibliography{custom}

\appendix

% \section{Conditional Statement and Code Complexity}
% \label{sec:appendix-cc}

\section{Instance of the APPS+ Dataset}
\label{sec:appendix-instance}

\begin{figure*}[t]
\centering
\includegraphics[width=0.98\textwidth]{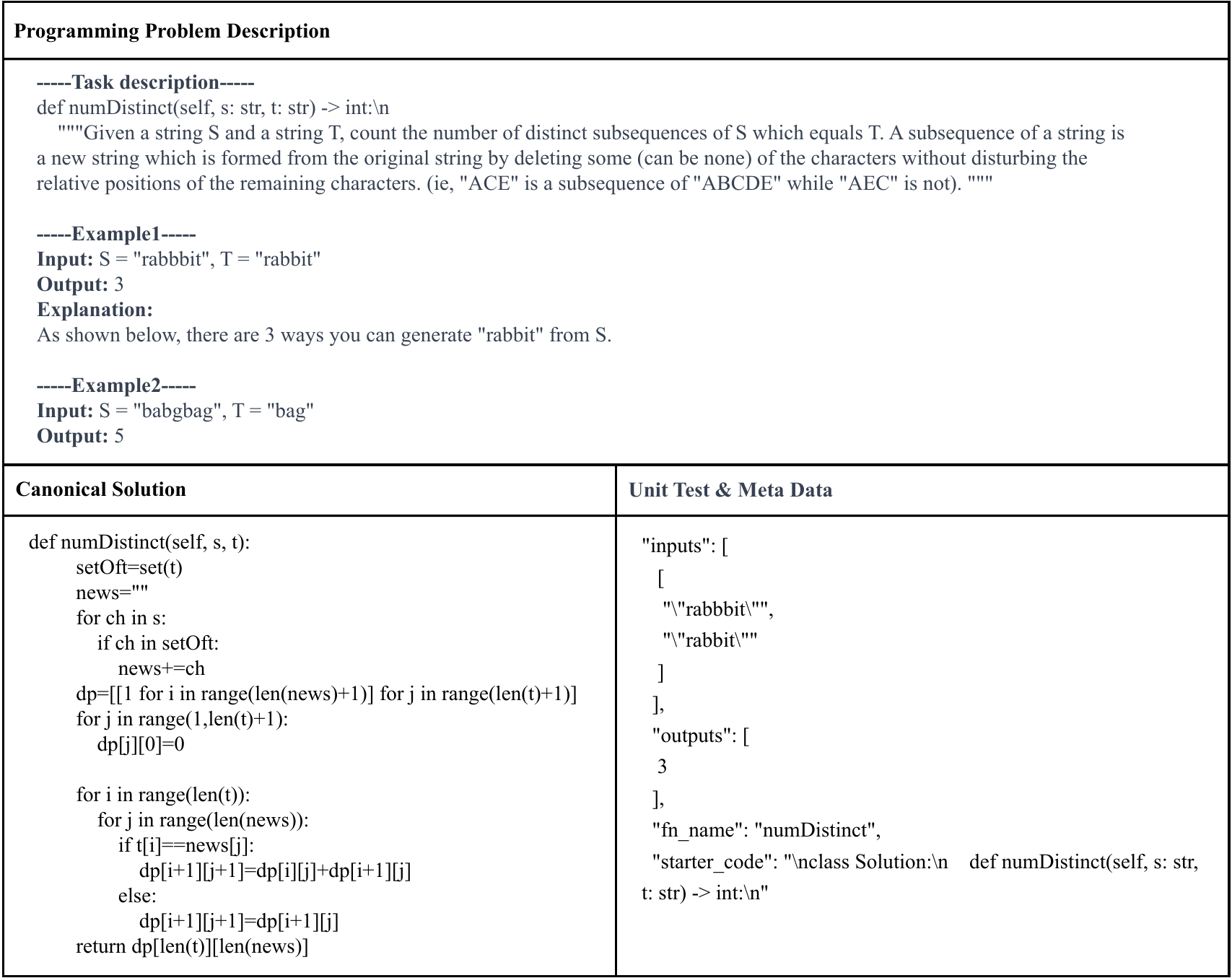}
\caption{An instance from our APPS+ dataset includes a human requirement (top), corresponding canonical code (bottom left), metadata, and example cases for unit testing to evaluate the generated code (bottom right).
We clean the APPS dataset \cite{apps} to provide a more rigorous evaluation and a foundation for training by RL in code generation.
}
\vspace{-1em}
\label{fig:fig-apps-plus-overview}
\end{figure*}

We present an example from our APPS+ dataset, as shown in Figure~\ref{fig:fig-apps-plus-overview}.

\section{Experiments Setup in Detail}
\label{sec:appendix-expriment}

In this section, we elaborate in detail on the baselines we compare and the implementation details of our method.

\subsection{Benchmarks}
\label{sec:appendix-exp-benchmarks}

\textbf{APPS+.}
We construct the new benchmark APPS+ by refining the popular benchmark APPS \cite{apps}.
APPS+ was categorized into three difficulty levels: Introductory (2,850), Interview (4,020), and Competition (586). 
The mean length of each problem is 255.3 words, and that of the code is 21.9 lines. 
On average, each instance is accompanied by three unit tests and includes a `conditional statement' attribute representing the start and end position of the statement in the canonical solution. 
We randomly selected about 25\% instances (700 Introductory, 1,000 Interview, and 140 Competition) for the validation dataset and another 25\% instances for the test dataset.

\textbf{MBPP.}
MBPP \cite{austin2021program} is a smaller but common Python code generation benchmark.
It contains 974 instances created by crowd-sourcing to an internal pool of crowd workers with basic Python knowledge.
The difficulty level of the problems in this dataset is introductory. Most problems are often conveyed in a single sentence of natural language, and each problem consists of a task description, code solution, and three automated test cases.
We evaluate LLMs in a zero-shot learning setting which is the same as previous studies \cite{le2022coderl, shojaee2023execution}.
In this setting, we fine-tune models only based on the APPS+ dataset and evaluate them on MBPP.

\textbf{HumanEval.}
HumanEval \cite{chen2021evaluating} is another extensively used benchmark for evaluating the ability of code generation.
It comprises 164 hand-written Python problems that test language comprehension, algorithmic thinking, and basic mathematics. 
The complexity of these problems is akin to that of simple software interview questions.
We also evaluate models on the HumanEval benchmark in a zero-shot learning setting.

\subsection{Baselines}
\label{sec:appendix-exp-baselines}
\textbf{StarCoder.}
StarCoder \cite{li2023starcoder} is a 15.5B parameter model trained on 80+ programming languages sourced from GitHub, encompassing one trillion tokens. It undergoes fine-tuning specifically for 35 billion Python tokens, enabling its proficiency across a diverse set of coding tasks. With an extended context length of 8K, StarCoder excels particularly in infilling capabilities.

\textbf{CodeLlama.}
CodeLlama \cite{roziere2023codellama} is a collection of pre-trained and fine-tuned generative text models ranging in scale from 7B to 34B parameters. CodeLlama comes in three variants: \textbf{CodeLlama}: base models designed for general code synthesis and understanding; \textbf{CodeLlama-Python}: designed specifically to handle the Python programming language; \textbf{CodeLlama-Instruct}: for instruction following and safer deployment.

\textbf{WizardCoder.}
WizardCoder \cite{luo2023wizardcoder} is fine-tuned by using a complicated dataset which is constructed by adapting the Evol-Instruct \cite{xu2023wizardlm} on code-related tasks, which is a further improvement of self-instruct method \cite{wang2022self}.
It has proven to be highly effective in code generation by fine-tuning more complex instruction data.

\textbf{DeepSeek-Coder.}
DeepSeek-Coder \cite{Guo2024DeepSeekCoderWT} demonstrates state-of-the-art performance among open-source code models across various programming languages.
It encompasses a collection of code language models from 1B to 33B trained from scratch. The training corpus for these models comprises an impressive 2 trillion tokens which is the combination of code and natural languages.
Each model is trained to utilize a window size of 16K, and a fill-in-the-blank task is incorporated into the training process, which enhances the models' capacity to facilitate code completion and infilling tasks.

\textbf{PPOCoder.}
PPOCoder \cite{shojaee2023execution} initially employs the Proximal Policy Optimization algorithm \cite{schulman2017proximal} for code generations. In addition, it integrates discrete compiler feedback with syntax and semantics matching scores between generated code and executable objectives which reduces the sparsity of the reward function, thereby providing better guidance for generating code that aligns more closely with the correct objectives.

\textbf{RLTF.}
RLTF \cite{liu2023rltf} features real-time data generation during the training process and multi-granularity unit test feedback. Except for the discrete compiler feedback, it penalizes specific sections in the code where errors occur through the error locations from the feedback of unit tests.

\section{The algorithm of CCCS and FGO}
\label{algorithm-stepcoder}
The full algorithm of StepCoder is detailed in Algorithm~\ref{alg:algorithm1}.

\begin{algorithm*}
\caption{StepCoder: Improve Code Generation with Reinforcement Learning from Compiler Feedback}
\label{alg:algorithm1}
\begin{algorithmic}[1]
\Require the train dataset $\mathcal{D} = \{(x_i, y_i, u_i, e_i), 1 \le i \le n\}$, the threshold value $\rho_t$ for curriculum training.
\Require the policy model $\pi_\theta$
% \REQUIRE the policy model $\pi_\theta$, the reference model $\pi^{\text{SFT}}$.
% Initialize sample metadata
\State Initialize the stride of curriculum $s =\lceil \frac{E_i}{\lceil \sqrt{E_i} \rceil} \rceil$ for each sample
\State Initialize the current curriculum $c = \lceil \sqrt{E_i} \rceil -1$ for each training sample
\State Initialize the pass rate $\rho = 0$ for each training sample
\While{TRUE}

% \STATE \textbf{\# Sampling trajectories phase, Stepper}
\State Initialize mini-batch $\mathcal{D}_s = \{\}$
\State Get latest policy model $\pi_\theta$
\State Sample a mini-batch of size $M$ from $\mathcal{D}$
\For {$i$ \text{in} $0$, $\cdots$, $M-1$} \Comment{Begin to sample the trajectories}
% \STATE $X_t = \{(x^i, y_w^i, y_l^i), 1 \le i \le n\} \gets$ SampleMiniBatch($\mathcal{D}$, $n$)
\State Calculate the start position $\text{pos} = s_i * c_i$ \Comment{CCCS}
\State Reorganize the given context $x^{'}_i = x_i + y_i\left[:\text{pos}\right]$
\State Sample trajectory $\hat{y_i} \gets \pi_\theta(.|x^{'}_i)$
\State Compute reward $r_i$ using Equation~\ref{eq:reward}
% \STATE \textbf{\# Finer}
\State Calculate unexecuted snippets' mask matrix $m_{ij} = [1\ \text{if} \ \hat{y}_{i}^{j} \ \text{is executed} \ \text{else} \ 0]$  \Comment{FGO}
\State Add $\{x^{'}_i, \hat{y_i}, u_i, r_i, s_i, c_i, m_i\}$ to mini-batch $\mathcal{D}_s$
\EndFor

% \STATE \textbf{\# Training phase}
% \FOR {$i = 0$, $\cdots$, $M-1$}
% \ENDFOR
\State $\theta \gets \mathcal{A}(\theta,\mathcal{D}_s)$ \Comment{Update the policy model by PPO algorithm}

\For {$i$ \text{in} $0$, $\cdots$, $M-1$} 
\If {$r_i = 1$} \Comment{Update pass rate using moving average}
\State $\rho_i = \alpha + (1-\alpha)*\rho_i$
\Else
\State $\rho_i = (1-\alpha)*\rho_i$
\EndIf
\If {$\rho_i > \rho_t$}  \Comment{Meet the update conditions, proceed to the next stage}
\State $\rho_i = 0$
\State $c_i = min(c_i -1,0)$
\EndIf
\EndFor
\EndWhile
\end{algorithmic}
\end{algorithm*}

\section{The prompts used for code generation}
\label{prompt}

For DeepSeek-Coder-Instruct \cite{Guo2024DeepSeekCoderWT}, we use the same prompt as the previous paper.
Moreover, DeepSeek-Coder-Instruct serves as the backbone model for PPOCoder \cite{shojaee2023execution}, RLTF \cite{liu2023rltf}, and our proposed StepCoder.
Consequently, we align the prompts for these RL-based approaches with the prompt of DeepSeek-Coder-Instruct to maintain consistency.
The prompt used for other models such as CodeLlama, WizardCoder and StarCoder is the same as in previous studies \cite{2023opencompass, luo2023wizardcoder, li2023starcoder, roziere2023codellama}.

The prompt used for DeepSeek-Coder-Instruct and LLMs based on it is as follows:

\noindent
\textit{You are an AI programming assistant, utilizing the Deepseek Coder model, developed by Deepseek Company, and you only answer questions related to computer science. } \\
\textit{For politically sensitive questions, security and privacy issues, and other non-computer science questions, you will refuse to answer.} \\
\textit{\#\#\# Instruction:} \\
\textit{write an algorithm in python:} \\
\textit{\{\textbf{Task description}\}} \\
\textit{\#\#\# Response:}

\end{document}